\documentclass[10pt]{article}

\usepackage[a4paper, margin=1in]{geometry}

\usepackage{amsmath, amssymb}

\usepackage{url}
\usepackage{graphicx}
\usepackage{xcolor}
\usepackage[normalem]{ulem}

\usepackage{authblk}

\usepackage[hidelinks]{hyperref}

\usepackage{listings}
\lstdefinelanguage{C}{
	keywords={break, case, catch, continue, debugger, default, delete, do, else, finally, for, function, if, in, instanceof, new, return, switch, throw, try, typeof, var, void, while, with, let, const, class, extends, super, export, import},
	morekeywords={async, await, yield},
	sensitive=true,
	morecomment=[l]{//},
	morecomment=[s]{/*}{*/},
	morestring=[b]",
	morestring=[b]',
	morestring=[b]`
}

\title{Exploring Heat Exchanges with the Calorimetry Simulator -- SimuF\'isica\textsuperscript{\textregistered}}

\author[1]{Marco P. M. de Souza\thanks{\href{mailto:marcopolo@unir.br}{marcopolo@unir.br}}}
\author[1,2]{Alex B. Siqueira}
\affil[1]{Departamento de F\'isica, Universidade Federal de Rond\^onia, Ji-Paran\'a, RO, Brazil.}
\affil[2]{Escola Estadual Angelina Franciscon Mazutti, Campos de J\'ulio, MT, Brazil.}

\begin{document}
	
	\maketitle
	
	\begin{abstract}
		This article presents the Calorimetry -- SimuF\'isica\textsuperscript{\textregistered} simulator, an interactive computational tool designed for teaching heat exchange processes. The simulator enables dynamic and audiovisual exploration of phenomena such as the heating of liquids and the establishment of thermal equilibrium between bodies at different temperatures. We describe its interface and the underlying physics, based on the equations of specific heat, latent heat of vaporization, and Newton's law of cooling. Two educational application examples are analyzed, in which the simulation results are compared with calculations typically performed at the high school and early undergraduate levels, highlighting the consistency between theory and simulation and the pedagogical potential of the tool.
	\end{abstract}
	
\section{Introduction}

The use of interactive computer simulations has become consolidated as an effective pedagogical strategy in science education, particularly in Physics. Several studies have shown that simulators can contribute to the understanding of abstract concepts, the visualization of phenomena invisible to the naked eye, and the development of students’ scientific reasoning and problem-solving skills \cite{Rutten2012, Jonassen2000, Banda2021}. In addition to promoting meaningful learning, these tools provide a controlled, accessible, and safe environment for experimentation, expanding the possibilities for conceptual exploration in both face-to-face and remote contexts.

The recognition of the relevance of simulations in school settings is also evidenced by official Brazilian documents such as the Base Nacional Comum Curricular (BNCC), which recommends the use of technological resources as facilitators of the teaching-learning process. In this context, initiatives such as the SimuFísica\textsuperscript{\textregistered} platform emerge, offering a set of interactive simulators developed to support Physics education at various levels.

SimuFísica\textsuperscript{\textregistered} \cite{Souza2024a, Souza2024b, Oliveira2024} is a multilingual, multiplatform educational software composed of dozens of simulators covering several topics in physics, such as mechanics, thermodynamics, electromagnetism, optics, and modern physics. The platform is freely available online\footnote{\url{https://simufisica.com/en}} and is also offered as a downloadable application for Windows, Android, iOS, and Linux. Key features include offline functionality, an intuitive interface, flexible configuration of physical parameters for each simulation, and the option to save simulations after user registration and login. As reported in recent studies, the platform has demonstrated potential to promote student engagement, autonomy in learning, and active knowledge construction.

In this article, we present the Calorimetry – SimuFísica\textsuperscript{\textregistered} application, which enables the simulation of processes involving liquid heating and thermal equilibrium. We begin by describing the interface and the physical foundations implemented in the simulation. Next, we explore two problem-solving examples applicable in the classroom: the heating of a liquid to the point of evaporation and the heat exchange between a metallic sphere and a liquid. Finally, we present our concluding remarks.

\section{The Calorimetry Simulator -- SimuF\'isica\textsuperscript{\textregistered}}

The Calorimetry simulator (Fig. 1), available on the SimuFísica\textsuperscript{\textregistered} platform, is an interactive tool designed for teaching and learning key concepts related to heat exchange. The simulation enables dynamic and audiovisual exploration of heating and thermal equilibrium processes, making it suitable for both high school and introductory undergraduate Physics courses.

\begin{figure}[h!]
	\centering
	\includegraphics[width=0.75\linewidth]{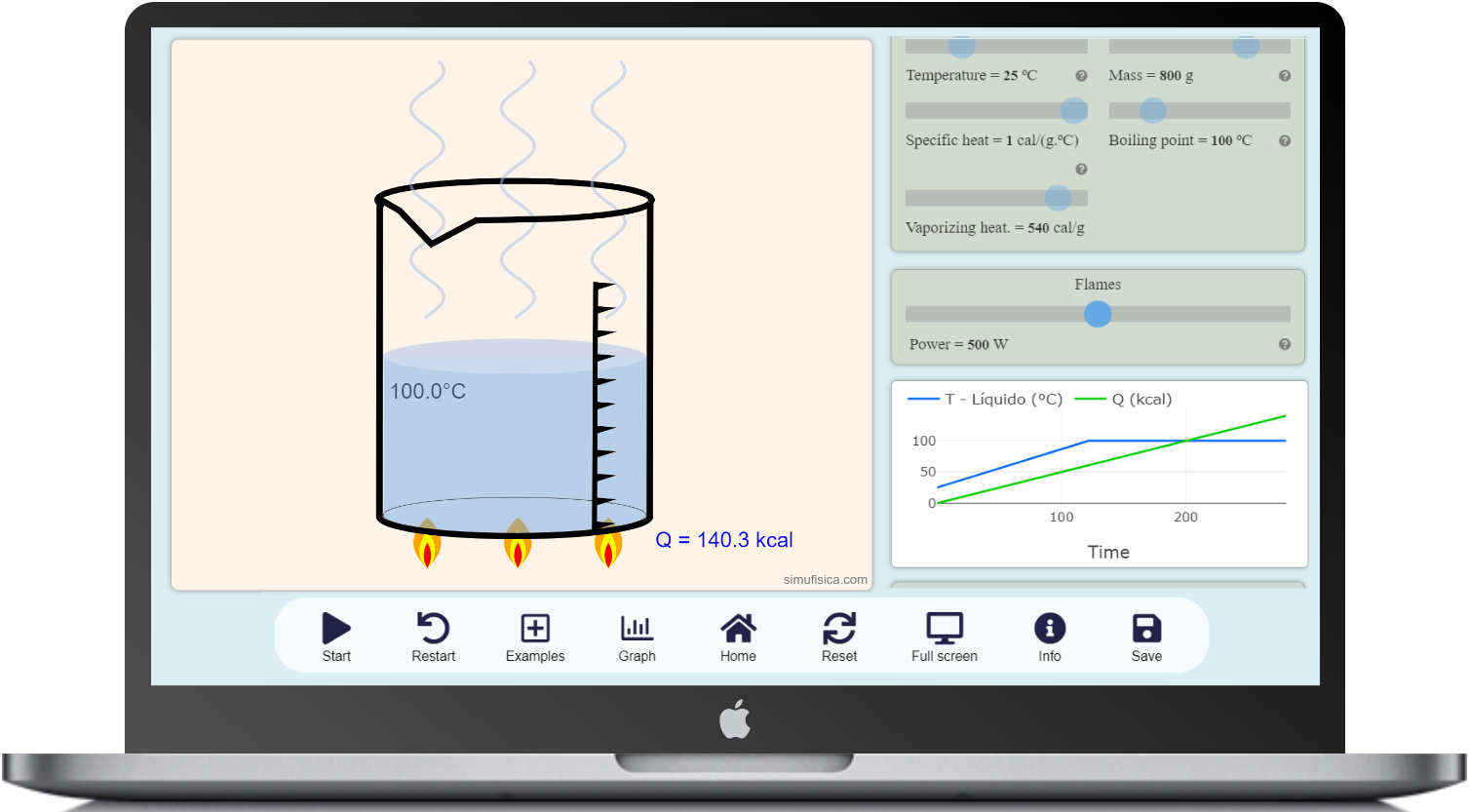}
	\caption{Calorimetry simulator from the SimuFísica\textsuperscript{\textregistered} platform, online desktop version configured in Heating mode. Access link: \url{https://simufisica.com/en/calorimetry/}}
	\label{fig:problema2}
\end{figure}

\subsection{Interface}

The simulator interface was designed to be intuitive and functional across different devices, from large-screen computers to smartphones, ensuring clarity in the representation of physical phenomena. The central visual elements are the beaker containing the liquid, the flame, and the metallic sphere (when applicable). Relevant physical parameters can be adjusted via numeric input fields located on side panels, as shown in Fig. 1.

There are two simulation modes: Heating and Thermal Equilibrium. In the Heating mode, the user observes the thermal energy transfer to a liquid being heated by a flame with adjustable power. Properties such as initial temperature, mass, specific heat, boiling point, and latent heat of vaporization can be configured. The simulation displays the heating process up to the boiling point and then the vaporization phase, automatically ending when the entire liquid has evaporated.

In Thermal Equilibrium mode, the scenario simulates the immersion of a heated metallic sphere into a colder liquid, with no external heat source. The simulation begins when the user releases the sphere into the container using drag-and-drop, either by mouse or touch, depending on the device. Besides the liquid's properties, it is also possible to configure the sphere’s initial temperature, mass, and specific heat. The simulation then tracks the heat exchange process until thermal equilibrium is reached, in accordance with the principle of energy conservation.

In both operating modes, the simulator displays in real time the heat absorbed by the liquid. A noteworthy feature is the presence of sound effects that signal specific events, such as the flame during heating or the impact of the sphere with the liquid and the container bottom, enhancing the sensory experience.

The lower toolbar includes two pedagogical tools: the Examples button, which provides predefined configurations (such as Ethanol Heating or Aluminum in Water), and the Graph button, which shows the time evolution of the liquid and sphere temperatures and the exchanged heat, enabling simultaneous and comparative analysis.

With simple commands such as the Start and Pause buttons, located in the bottom toolbar, users have direct control over the simulation and can switch between qualitative and quantitative analysis. Additionally, registered users can use the Save button to store simulations with their respective parameters for future access, facilitating fast setup for classroom activities. These controls, along with others such as the Home button, compose the standard interface used across various SimuFísica simulators --- see references \cite{Souza2024a, Souza2024b, Oliveira2024, Souza2025} for more details.

\subsection*{B. Fundamentals}

In Heating mode, the simulator models the temperature rise of a liquid heated by a flame with adjustable, but constant, power $P$. For temperatures below the boiling point, the heat $Q$ absorbed is proportional to the temperature increase $\Delta T$, as given by Eq.~(\ref{eq1}):

\begin{equation}\label{eq1}
	Q = m_0 c \Delta T
\end{equation}

\noindent where $m_0$ is the mass of the liquid and $c$ is its specific heat. The amount of heat provided by the flame during a time interval $\Delta t$ is given by Eq.~(\ref{eq2}):

\begin{equation}\label{eq2}
	Q = P \Delta t
\end{equation}

\noindent Assuming that all the energy from the flame is absorbed by the liquid — neglecting thermal losses and the heat capacity of the container — we combine Eqs.~(\ref{eq1}) and~(\ref{eq2}) to obtain the temperature of the liquid as a function of time:

\begin{equation}\label{eq3}
	T(t) = T_0 + \frac{P t}{m_0 c}, \quad \text{for } t \le t_E
\end{equation}

\noindent where $T_0$ is the initial temperature. The time $t_E$ when the liquid reaches the boiling point $T_E$ is given by Eq.~(\ref{eq4}):

\begin{equation}\label{eq4}
	t_E = \frac{m_0 c (T_E - T_0)}{P}
\end{equation}

\noindent After $t_E$, the heat is used to vaporize the liquid. The heat required to vaporize a mass $m$ is given by Eq.~(\ref{eq5}):

\begin{equation}\label{eq5}
	Q = m L
\end{equation}

\noindent where $L$ is the latent heat of vaporization. Combining Eqs.~(\ref{eq2}) and~(\ref{eq5}), the mass evaporated over time for $t > t_E$ is:

\begin{equation}\label{eq6}
	m_{\text{vap}}(t) = \frac{P (t - t_E)}{L}
\end{equation}

\noindent Equations~(\ref{eq1}),~(\ref{eq3}) and~(\ref{eq6}) form the basis of the physical model used in the simulator for Heating mode.

In Thermal Equilibrium mode, while the liquid temperature remains below the boiling point, the thermal evolution of the system is governed by Newton’s law of cooling, resulting in the system of ordinary differential equations~(\ref{eq7}):

\begin{subequations}\label{eq7}
	\begin{align}
		c_{\text{liq}} \frac{dT_{\text{liq}}}{dt} &= -k (T_{\text{liq}} - T_{\text{esf}}) \label{eq7a} \\
		c_{\text{esf}} \frac{dT_{\text{esf}}}{dt} &= \phantom{-}k (T_{\text{liq}} - T_{\text{esf}}) \label{eq7b}
	\end{align}
\end{subequations}

\noindent where $c_{\text{liq}}$ and $c_{\text{esf}}$ are the specific heats of the liquid and sphere, respectively, and $k$ is a constant related to the heat transfer rate, adopted as $1\, \text{cal}/(\text{g} \cdot ^\circ\text{C} \cdot \text{s})$ in the simulator.

The solutions to system~(\ref{eq7}), with initial conditions $T_{\text{liq}}(0) = T_0^{\text{liq}}$ and $T_{\text{esf}}(0) = T_0^{\text{esf}}$, are given in Eqs.~(\ref{eq8}):

\begin{subequations}\label{eq8}
	\begin{align}
		T_{\text{liq}}(t) &= \frac{c_{\text{esf}} T_0^{\text{esf}} + c_{\text{liq}} T_0^{\text{liq}}}{c_{\text{esf}} + c_{\text{liq}}} + \frac{c_{\text{esf}} (T_0^{\text{liq}} - T_0^{\text{esf}})}{c_{\text{esf}} + c_{\text{liq}}} \exp\left( -\frac{k (c_{\text{esf}} + c_{\text{liq}})}{c_{\text{esf}} c_{\text{liq}}} t \right) \label{eq8a} \\
		T_{\text{esf}}(t) &= \frac{c_{\text{esf}} T_0^{\text{esf}} + c_{\text{liq}} T_0^{\text{liq}}}{c_{\text{esf}} + c_{\text{liq}}} - \frac{c_{\text{liq}} (T_0^{\text{liq}} - T_0^{\text{esf}})}{c_{\text{esf}} + c_{\text{liq}}} \exp\left( -\frac{k (c_{\text{esf}} + c_{\text{liq}})}{c_{\text{esf}} c_{\text{liq}}} t \right) \label{eq8b}
	\end{align}
\end{subequations}

\noindent When the liquid reaches its boiling point $T_E^{\text{liq}}$, its temperature remains constant and the energy transferred from the sphere is used exclusively for vaporization. The model implemented in the simulator adopts the following equations~(\ref{eq9}):

\begin{subequations}\label{eq9}
	\begin{align}
		T_{\text{liq}}(t) &= T_E^{\text{liq}} \label{eq9a} \\
		T_{\text{esf}}(t) &= T_E^{\text{liq}} + (T_0^{\text{esf}} - T_E^{\text{liq}}) \exp\left( -\frac{k}{c_{\text{esf}}} (t - t_E^{\text{liq}}) \right) \label{eq9b} \\
		m_{\text{vap}}(t) &= \frac{ \left| m_{\text{esf}} c_{\text{esf}} (T_{\text{esf}} - T_{\text{liq}}) \right| - m_{\text{liq}} c_{\text{liq}} (T_E^{\text{liq}} - T_{\text{liq}}) }{L} \label{eq9c}
	\end{align}
\end{subequations}

\noindent where $m_{\text{esf}}$ and $m_{\text{liq}}$ are the masses of the sphere and the initial liquid, respectively, and $m_{\text{vap}}$ is the vaporized mass.

Equations~(\ref{eq8}) and~(\ref{eq9}) describe the system dynamics in Thermal Equilibrium mode and allow the simulator to faithfully reproduce the heat exchange process, with and without phase change.

\section{Application Examples}

We present in this section two examples of educational activities using the \textit{Calorimetry} simulator, aiming to illustrate its use in the classroom to foster understanding of fundamental concepts related to heat exchange and phase changes.

\subsection*{Problem 1}

A container holds 500~g of water at 20\,°C. This water is heated by a flame that transfers 148~kcal of heat over time. Neglect the heat capacity of the container. The specific heat of water is 1~cal/(g.°C) and the latent heat of vaporization is 540~cal/g.

\begin{itemize}
	\item[(a)] What is the final temperature of the water?
	\item[(b)] Did any water evaporate? If so, what was the evaporated mass?
\end{itemize}

\textbf{Solution.} To solve item (a), we use the sensible heat equation:
\begin{equation}\label{eq10}
	Q = mc\Delta T
\end{equation}

\noindent where $Q = 148$~kcal, $m = 500$~g~$= 0.5$~kg, and $c = 1$~kcal/(kg.°C). Solving Eq.~\eqref{eq10} for $\Delta T$:
\begin{equation}\label{eq11}
	\Delta T = \frac{Q}{mc} = \frac{148}{0.5 \times 1} = 296\,^\circ\text{C}
\end{equation}

\noindent Since the water starts at 20\,°C, the final temperature would be:
\begin{equation}\label{eq12}
	T_f = 20 + 296 = 316\,^\circ\text{C}
\end{equation}

\noindent (This can be confirmed in the simulator by artificially increasing the boiling point above 316\,°C.) However, under atmospheric pressure, water boils at 100\,°C, so it cannot exceed this temperature. The excess energy causes evaporation. The heat needed to bring the water from 20\,°C to 100\,°C is:
\begin{equation}\label{eq13a}
	Q_1 = mc(100 - 20) = 0.5 \times 1 \times 80 = 40\,\text{kcal}
\end{equation}

\noindent Thus, the remaining energy for the phase change is:
\begin{equation}\label{eq13b}
	Q_2 = 148 - 40 = 108\,\text{kcal}
\end{equation}

\noindent The evaporated mass is calculated using:
\begin{equation}\label{eq14}
	m_{\text{vap}} = \frac{Q_2}{L} = \frac{108}{540} = 200\,\text{g}
\end{equation}

\noindent Therefore, the final temperature is 100\,°C and 200\,g of water evaporate, leaving 300\,g in the beaker.

\begin{figure}[h!]
	\centering
	\includegraphics[width=0.75\linewidth]{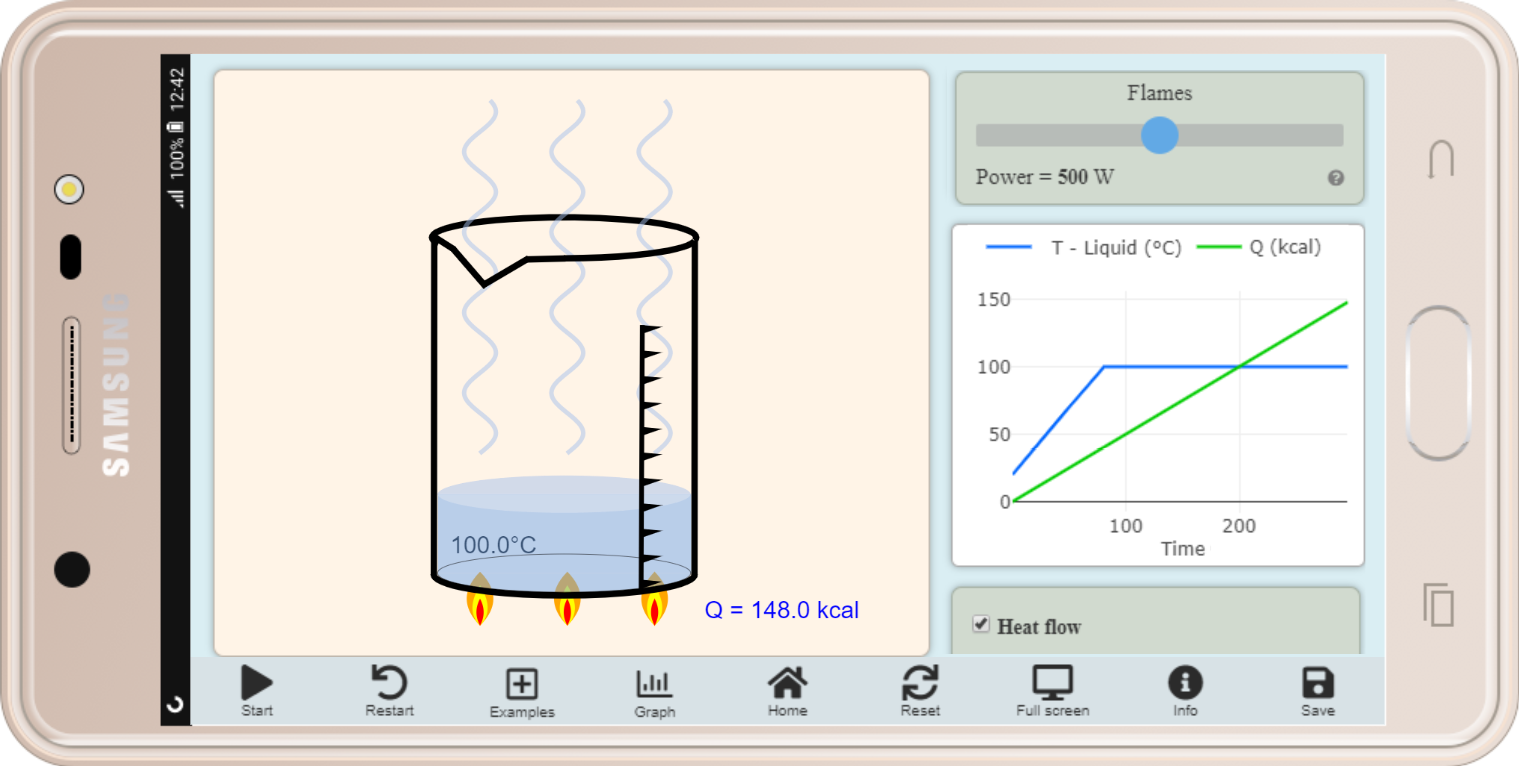}
	\caption{Simulator configured for Problem 1, in Heating mode and online smartphone version, showing evaporation after reaching 100\,°C. The graph displays temperature and absorbed heat versus time. Link to this configuration: \url{https://simufisica.com/5OX84u}.}
	\label{fig:problema1}
\end{figure}

\textbf{Comparison with the simulator.} Using the specified parameters (Heating mode, 500\,g of water at 20\,°C, specific heat 1~cal/g.°C, latent heat 540~cal/g), the simulator shows a linear temperature increase up to 100\,°C, consistent with Eq.~\eqref{eq3}. After that, the temperature curve flattens, indicating the boiling phase (see Fig.~\ref{fig:problema1}).

Simultaneously, the value of absorbed heat continues to increase, confirming vaporization. The water level (see the beaker scale in Fig.~\ref{fig:problema1}) begins to drop after boiling starts. Once 148~kcal are absorbed, the level drops from 500\,g to 300\,g, matching the theoretical result (each 100\,g corresponds to a scale mark). This example demonstrates how the simulator supports both qualitative and quantitative understanding of sensible heat, latent heat, and phase changes, enabling comparative analysis between theory and simulation and enhancing graph interpretation skills.

\subsection*{Problem 2}

A steel sphere of 200\,g, heated to 600\,°C, is placed in a container holding 700\,mL of water at 25\,°C. Neglect the container’s heat capacity. No phase change occurs.

\begin{itemize}
	\item[(a)] Find the final equilibrium temperature.
	\item[(b)] Calculate the amount of heat transferred from the sphere to the water.
\end{itemize}

Given: Specific heat of steel: 0.1~cal/g.°C; specific heat of water: 1~cal/g.°C; density of water: 1~g/cm$^3$.

\textbf{Solution.} The water mass is:
\begin{equation}\label{eq15}
	m_{\text{water}} = 700\,\text{g}
\end{equation}

Let $T_f$ be the final temperature. Since there is no heat loss or phase change:
\begin{equation}\label{eq16}
	m_{\text{steel}} c_{\text{steel}} (T_{\text{steel}} - T_f) = m_{\text{water}} c_{\text{water}} (T_f - T_{\text{water}})
\end{equation}

Substituting the values into Eq.~\eqref{eq16}:
\begin{equation}\label{eq17}
	200 \times 0.1 \times (600 - T_f) = 700 \times (T_f - 25)
\end{equation}

Expanding and rearranging:
\begin{equation}\label{eq18}
	12000 - 20T_f = 700T_f - 1750
\end{equation}

Solving for $T_f$:
\begin{equation}\label{eq19}
	T_f \approx 40.97\,^\circ\text{C}
\end{equation}

Now, using Eq.~\eqref{eq10}, the heat transferred from the sphere is:
\begin{equation}\label{eq20}
	Q = 200 \times 0.1 \times (600 - 40.97) \approx 11.18\,\text{kcal}
\end{equation}

\textbf{Comparison with the simulator.} When configured in Thermal Equilibrium mode, with the sphere at 600\,°C and water at 25\,°C, the simulator shows the sphere cooling and the water heating until both reach about 41\,°C, as shown in Fig.~\ref{fig:problema2}.

\begin{figure}[h!]
	\centering
	\includegraphics[width=0.75\linewidth]{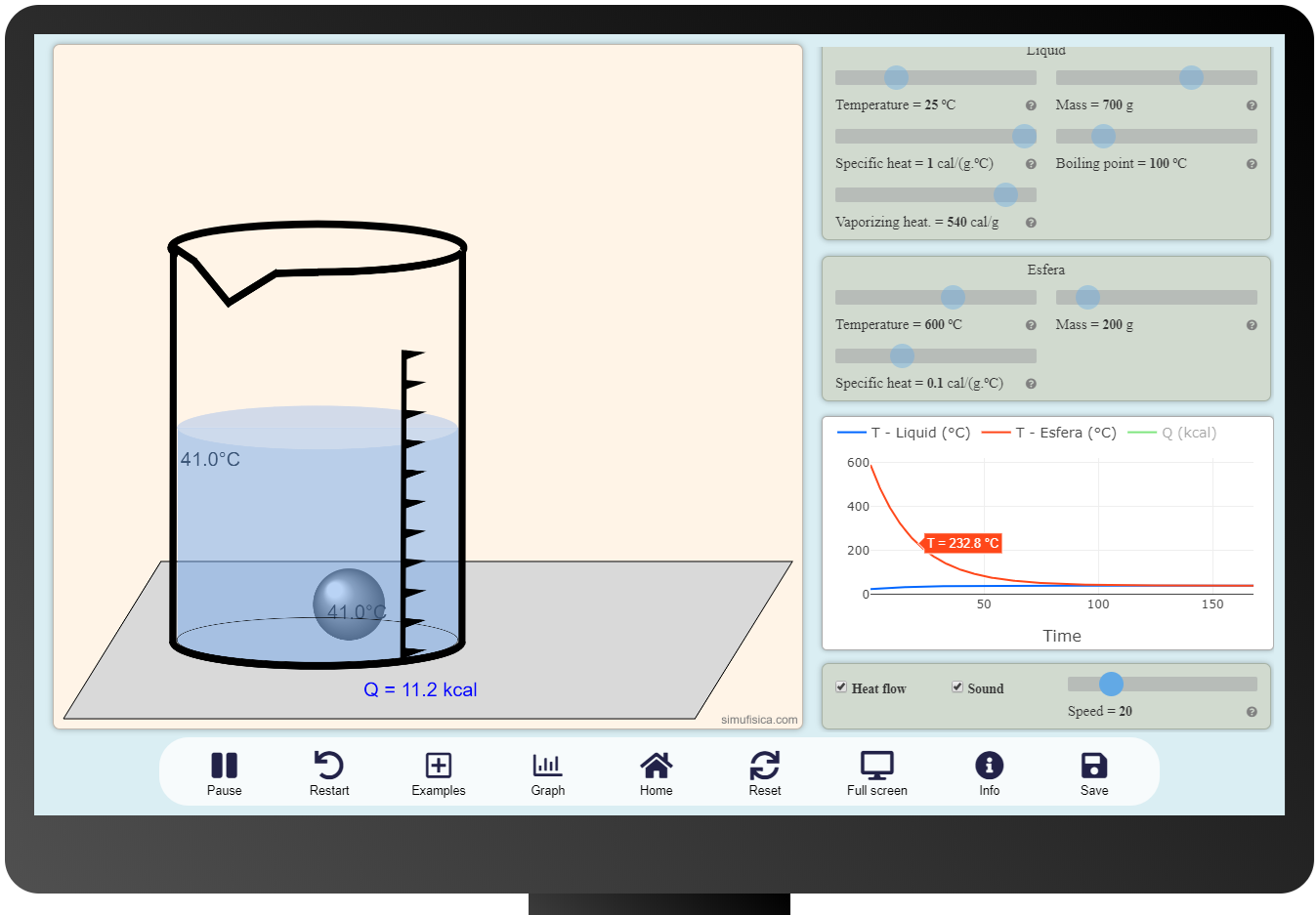}
	\caption{Simulator configured for Problem 2, in Thermal Equilibrium mode (desktop version). Final temperature is 41\,°C, with a total transferred heat of 11.2\,kcal. The graph shows temperatures of water and sphere versus time. Link to this configuration: \url{https://simufisica.com/9G4sxv}.}
	\label{fig:problema2}
\end{figure}

The water's temperature rises smoothly, while the sphere’s temperature drops rapidly. This behavior illustrates how different specific heats affect temperature variation. The total absorbed heat shown by the simulator is 11.2\,kcal, in agreement with the theoretical result from Eq.~\eqref{eq20}. This reinforces the usefulness of the simulator in visualizing thermal equilibrium processes, providing a clear and quantitative graphical representation of the phenomenon.

\section{Final Considerations}

This work presented the \textit{Calorimetry – SimuFísica\textsuperscript{\textregistered}} simulator, highlighting its potential as a support tool for teaching fundamental concepts related to heat exchange. By enabling the simulation of heating and thermal equilibrium processes, the simulator offers students an interactive, intuitive, and visually rich experience that facilitates the construction of meaning around topics that may pose challenges in Physics education.

The implementation of the physical models behind the simulation, based on the equations for sensible heat, latent heat, and Newton's law of cooling, ensures the fidelity of the representations and allows students to compare theoretical results with data produced in the simulation. The comparison between the analytical calculations and the graphical results obtained in the two proposed problems—one involving heating with phase change, and the other involving thermal equilibrium without phase change—demonstrated the simulator's consistency with classical mathematical models and its usefulness for developing skills such as graph interpretation, experimental estimation, and understanding of energy conservation.

Moreover, the inclusion of features such as sound effects and predefined examples enhances student engagement and accessibility, allowing educators to use the tool both as a complement to conceptual instruction and as a resource for inquiry-based activities or computational simulations.

As a future perspective, we emphasize the importance of conducting systematic studies on the pedagogical impact of the \textit{Calorimetry} simulator in various educational contexts, as well as exploring the integration of interactive activities based on this simulator with active methodologies, such as problem-based learning and hybrid teaching (part in-person, part remote). Another possibility would be to compare the real heating of liquids in the laboratory with the simulator's results. Finally, enhancing the graphical interface and including more configurable parameters also represent opportunities for improving the tool, further consolidating its role as a valuable resource for teaching Physics.

\section*{Acknowledgments}

This work was supported by the Conselho Nacional de Desenvolvimento Científico e Tecnológico (CNPq, Grant 304017/2022-1), the Fundação Rondônia de Amparo ao Desenvolvimento das Ações Científicas e Tecnológicas e à Pesquisa do Estado de Rondônia (FAPERO, Grant 36214.577.20546.20102023), and the Universidade Federal de Rondônia (UNIR, Grant 23118.006316/2024-79).


\begin{thebibliography}{99}
	
	\bibitem{Rutten2012} N. Rutten, W. R. van Joolingen, and J. T. van der Veen, \textit{The learning effects of computer simulations in science education}, Comput. Educ. \textbf{58}, 136 (2012). doi:10.1016/j.compedu.2011.07.017
	
	\bibitem{Jonassen2000} D. H. Jonassen, \textit{Computers as Mindtools for Schools} (2nd ed.), Merrill, New Jersey, USA, 2000.
	
	\bibitem{Banda2021} H. J. Banda and J. Nzabahimana, \textit{Effect of integrating physics education technology simulations on students’ conceptual understanding in physics: A review of literature}, Phys. Rev. Phys. Educ. Res. \textbf{17}, 023108 (2021). doi:10.1103/PhysRevPhysEducRes.17.023108
	
	\bibitem{Souza2024a} M. P. M. de Souza, C. M. Oliveira, and R. P. P. Araújo, \textit{O simulador conservação de energia mecânica – SimuFísica\textsuperscript{\textregistered}}, A Fís. na Escola \textbf{22}, 240173 (2024). doi:10.59727/fne.v22i1.173. English version (arXiv): \url{https://doi.org/10.48550/arXiv.2410.17951}
	
	\bibitem{Souza2024b} M. P. M. de Souza, S. P. Oliveira, and V. L. Luiz, \textit{Motor elétrico – SimuFísica\textsuperscript{\textregistered}: um aplicativo para o ensino de eletromagnetismo}, Rev. Bras. Ens. Fís. \textbf{46}, e20230219 (2024). doi:10.1590/1806-9126-RBEF-2023-0219. English version (arXiv): \url{https://doi.org/10.48550/arXiv.2410.18721}
	
	\bibitem{Oliveira2024} C. M. Oliveira, \textit{Ensino da temática energia com três aplicativos da plataforma SimuFísica no Novo Ensino Médio}, Master's thesis, Universidade Federal de Rondônia -- UNIR, 2024.
	
	\bibitem{Souza2025} M. P. M. de Souza, G. H. H. Pavão, A. A. C. de Almeida, and S. S. Vianna, \textit{Bloch Equation Generator -- SimuFísica}, arXiv:2506.01108 [quant-ph]. \url{https://doi.org/10.48550/arXiv.2506.01108}
	
	\bibitem{Silva2003} W. P. Silva, J. W. Precker, C. M. D. P. S. Silva, D. D. P. S. Silva, and C. D. P. S. Silva, \textit{Medida de calor específico e lei de resfriamento de Newton: um refinamento na análise dos dados experimentais}, Rev. Bras. Ens. Fís. \textbf{25}(4), 392 (2003). doi:10.1590/S0102-47442003000400015
	
\end{thebibliography}
\end{document}